\newcommand{\be}{\begin{equation}}
\newcommand{\ee}{\end{equation}}
\newcommand{\bea}{\begin{eqnarray}}
\newcommand{\eea}{\end{eqnarray}}

\documentstyle[11pt]{article}
\begin{document}
\leftline{\scriptsize
{``Recent Developments in Gravitation and Mathematical Physics",
Proc. of the 2nd Mexican School on
Gravitation and Mathematical Physics,}}
\leftline{\scriptsize
{Eds. A. Garc\'{\i}a, C. Laemmerzahl, A. Mac\'{\i}as, T. Matos, D. Nu\~nez;
Science Network Publishing, Konstanz (1997) [ISBN 3-9805735-0-8]}}
\leftline{\scriptsize{Home page at http://kaluza.physik.uni-konstanz.de/2MS}}
\leftline{\scriptsize{Tlaxcala, December 1-6, 1996; speaker author Socorro}}

\begin{center}{\Large{\bf Supersymmetric double Darboux method in
quantum cosmology}}\\
\vskip 3mm
J. Socorro$^{\dagger}$\footnote {e-mail: socorro@ifug4.ugto.mx} and
H.C. Rosu$^{{\dagger},*}$\footnote {e-mail: rosu@ifug.ugto.mx}  \\

$^{\dagger}$
Instituto de F\'{\i}sica de la Universidad de Guanajuato, Apdo Postal
E-143, Le\'on, Gto, M\'exico\\
$^*$
Institute of Gravitation and Space Sciences, P.O. Box MG-6,
Magurele-Bucharest, Romania

\vskip 0.3cm

{\bf Abstract:}
We briefly present the supersymmetric double Darboux
method and next apply it to the continuum
of the quantum Taub cosmological model as a toy model in order to generate 
a one-parameter family of bosonic Taub potentials and the
corresponding wavefunctions.
\end{center}

\bigskip
Recently, extensive work has been devoted to generating quasi-isospectral 
potentials by the methods of supersymmetric quantum mechanics (SUSYQM) 
\cite{Wi}. This work starts from the one dimensional (1D) Schr\"odinger 
equation with a given potential (called ``bosonic" potential $\rm V_-$), 
whose ``ground-state" (nodeless) wave function is known too. The method
allows one to 
generate families of new potentials, which may look quite different form the 
original one, but have the same discrete spectrum up to the ground state.
SUSYQM  is in fact based on a procedure already known to Darboux
\cite{Da}. We shall use an extension of SUSYQM formalism based on the general
Riccati equation and on a state ${\cal U}_0(x)$ of a
potential $\rm V_-(x)$,
by which one can generate a one-parameter family of bosonic potentials
$\rm \hat V(x;\lambda)$ including $\rm V_-(x)$ \cite{KS}.
In SUSYQM, the labeling, real parameter $\lambda$ should be
positive for radial problems and should not be in the interval $[-1,0]$ for
the full line problems, but in quantum cosmology this constraints can be
relaxed.

We start with the 1D Hamiltonian
\be
\rm H= - \frac{d^2}{dx^2 } + V_-(x),
\label {schro}
\ee
which can be factored as $\rm H= A^+ A$ by means  of the operators A and 
$\rm A^+$ defined as
\be
\rm A= \frac{d}{dx} + W(x), \qquad\qquad A^+ =- \frac{d}{dx} + W(x).
\ee
In this way, one can get $\rm V_-(x)= W^2(x) - \frac{dW(x)}{dx}$, where
$\rm W(x)$ is called the superpotential. If ${\cal U}_0(x)$ is the ground state
solution of the 1D Schr\"odinger equation, the superpotential is given by
\be
\rm W(x)= -\frac{d}{dx} \lbrack ln {\cal U}_0(x) \rbrack.
\label {particular}
\ee
 In the supersymmetric scheme there exists another potential, called
``fermionic" potential, which is the partner of the bosonic one, that can be 
written as 
\be
\rm V_+(x)= W^2 + \frac{dW}{dx} = V_- - 2 \frac{d^2}{dx^2}
\lbrack ln {\cal U}_0(x) \rbrack .
\ee
It corresponds to the  supersymmetric partner Hamiltonian  
$\rm H'=A A^+ = -\frac{d^2}{dx^2}+ V_+(x)$.

If the potential $\rm V_-(x)$ has eigenfunctions ${\cal U}_n(x)$ at
energies $\rm E_n$, then the SUSY partner potential $\rm V_+(x)$ has the
same energy eigenvalues as $\rm V_-(x)$ with eigenfunctions 
$\rm A {\cal U}_n(x)$, except that there is no ground state at E=0
since $\rm A {\cal U}_0(x)=0$. This is the standard procedure for deleting the
ground state and obtaining $\rm V_+(x)$.
Now, suppose one asks for the most general superpotential
such that $\rm V_+(x)= \hat W^2 + \frac{d \hat W}{dx}$. It is easy to see that
one particular solution to this equation is $\rm W_p= W(x)$, where W(x) is 
given by Eq.(\ref {particular}). One is led to consider the following
Riccati equation $\rm \hat W^2 + \frac{d \hat W}{dx}=W^2_p +\frac{d W_p}{dx}$,
whose general solution can be written down as 
$\rm \hat W(x)= W_p(x) + \frac{1}{v(x)}$, where $\rm v(x)$ is an unknown 
function. Using this ansatz, one obtains for the function $\rm v(x)$ the 
following Bernoulli equation
\be
\rm \frac{dv(x)}{dx} - 2 \, v(x)\, W_p(x) = 1,
\ee
that has the following solution
\be
\rm v(x)= \frac{{\cal I}_0(x)+ \lambda}{{\cal U}_{0}^{2}(x)},
\ee
where ${\cal I}_0(x)= \int^x \, {\cal U}_0^2(y)\, dy $. Thus, $\rm \hat W(x)$
can be written as follows
\bea
\rm \hat W(x)&=& W_p(x) + \frac{d}{dx} \lbrack ln ({\cal I}_0(x) + \lambda) 
\rbrack \nonumber\\
 &=&\rm - \frac{d}{dx} \lbrack ln (\frac{{\cal U}_0(x)}{{\cal I}_0(x) + 
\lambda}) 
\rbrack. 
\label {general}
\eea
Finally, one gets easily the family of ``bosonic"
potentials 
\bea
\rm \hat V(x;\lambda) &=& \rm \hat W^2(x;\lambda) - 
\frac{d \hat W(x;\lambda)}{dx} \nonumber\\
&=& \rm V_-(x) - 2 \frac{d^2}{dx^2} \lbrack ln({\cal I}_0(x) + \lambda) 
\rbrack \nonumber\\
&=& \rm V_-(x) - \frac{4 {\cal U}_0(x) {\cal U}_0^\prime (x)}{{\cal I}_0(x) 
+ \lambda} 
+ \frac{2 {\cal U}_0^4(x)}{({\cal I}_0(x) + \lambda)^2}.
\eea
All $\rm \hat V(x;\lambda)$ have the same supersymmetric partner potential
$\rm V_+(x)$ obtained by deleting the ground state.
 
On the other hand, by inspection of Eqs. (\ref {particular}) and 
(\ref {general}) we can obtain the new family wave function
for the potentials $\rm \hat V(x;\lambda)$
\be
\rm \hat {\cal U}(x;\lambda)= f(\lambda)
\frac{{\cal U} _0(x)}{{\cal I}_0(x) + \lambda}.
\ee

In SUSYQM, some authors have argued that in order to have normalizable
wave functions, the 
function $\rm f(\lambda)$ must be of the following form
$\rm f(\lambda)= \sqrt{\lambda(\lambda +1)}$. A connection with
other isospectral methods has been found, by noticing that
the limiting values -1 and 0 for the parameter $\lambda$ lead to the
Abraham-Moses procedure
\cite {AbMo} and Pursey's \cite {Pu} one, respectively.
On the other hand, in quantum cosmology we shall work with $f(\lambda)=1$,
i.e., with nonnormalizable solutions.

The supersymmetric double Darboux method can be applied to any one-dimensional
system, whose dynamics is dictated by a Schr\"odinger-like equation.
Here we shall apply it by employing a wave function belonging to the
continuum of the quantum minisuperspace
Taub model, which is one of the simplest models in quantum cosmology.
It has been 
studied in some detail by Ryan and collaborators \cite {Ry}, who found that 
it is separable. Indeed, the Taub Wheeler-DeWitt equation is
\be
\rm \frac{\partial ^2\Psi}{\partial \alpha ^2}-
\frac{\partial ^2\Psi}{\partial \beta ^2}+e^{4\alpha}V(\beta)\Psi=0,
\label{WDW}
\ee
where $\rm V(\beta)=\frac{1}{3}(e^{-8\beta}-4e^{-2\beta})$. Eq.(\ref{WDW})
can be separated in the variables
$x_1=4\alpha-8\beta$ and $x_2=4\alpha -2\beta$.
Thus, one gets two independent 1D problems for which the supersymmetric
procedures are standard practice. The two equations are as follows
\bea
-\frac{\partial ^2 u_1}{\partial x_1^2}+\frac{1}{144}e^{x_1}u_1 &=&
\frac{\omega ^2}{4}u_1, \\
-\frac{\partial ^2 u_2}{\partial x_2^2}+\frac{1}{9}e^{x_2}u_2 &=&
\omega ^2 u_2 .
\eea
The quantity $\omega$ is mathematically the separation constant, which
physically is related to the wavenumber of a positive energy level.

Mart\'{\i}nez and Ryan have considered a wavepacket solution made of
wavefunctions $\Psi$ having
the form of a product of modified Bessel functions of imaginary order.
We shall slightly modify their $\Psi$ as follows
\be
\rm \Psi\equiv u_1u_2=K_{i\omega}(\frac{1}{6}e^{x_1/2})
[L_{2i\omega}(\frac{2}{3}e^{x_2/2})+
K_{2i\omega}(\frac{2}{3}e^{x_2/2})]
\label{solu}
\ee
since, according to Dunster \cite{Du}, the L function defined as
\be
\rm L_{2i\omega}=\frac{\pi i}{2{\rm sinh}
(2\omega \pi)}(I_{2i\omega}+I_{-2i\omega})
\ee
contrary to the $\rm I_{2i\omega}$ function used by Mart\'{\i}nez and Ryan,
being real on the real axis is a better companion for the K function of
imaginary order, which is known to be real on the real axis.

In the following, we shall make the double Darboux construction on the 
base of a $\Psi$ wavefunction of the type given in Eq. (\ref{solu}).

The family of bosonic potentials for the $x_1$ variable will be
\be
\rm V_1(x_1;\lambda _1)=\frac{1}{144}e^{x_1}-2\frac{d^2}{dx_1^2}
\ln[\lambda _1 +{\cal I} (x_1)]
\label {uno}
\ee
and for the $x_2$ one
\be
\rm V_2(x_2;\lambda _2)=\frac{1}{9}e^{x_2}-2\frac{d^2}{dx_2^2}
\ln[\lambda _2 +{\cal I} (x_2)]
\label {dos}
\ee
where the integrals are
\be
\rm {\cal I} (x_1)=\int_{-\infty}^{x_1}K_{i\omega}^{2}(\frac{1}{6}e^{y/2})dy,
\qquad\qquad
{\cal I} (x_2)=\int_{-\infty}^{x_2}[L_{2i\omega}(\frac{2}{3}e^{y/2})+
K_{2i\omega}(\frac{2}{3}e^{y/2})]^2 dy~.
\ee
The total wavefunction of the bosonic family of Taub potentials
has the following form
\be
\rm \hat \Psi ^{T}(x_1, x_2;\lambda _1, \lambda _2)\equiv
\hat u_{1}\hat u_{2}= \frac{
K_{i\omega}(\frac{1}{6}e^{\frac{1}{2}x_1})}{[{\cal I} (x_1)+\lambda _1]}
\frac{(L_{2i\omega}(\frac{2}{3}e^{\frac{1}{2}x_2})+
K_{2i\omega}(\frac{2}{3}e^{\frac{1}{2}x_2}))}
{[{\cal I} (x_2)+\lambda _2]} .
\ee

In conclusion, we used
the SUSYQM double Darboux method to introduce
a one-parameter family of ``quantum" Taub potentials
having a different set of wavefunctions in the continuum region of the
spectrum with respect to the original potential. For more details see
ref. {\cite{Ta}.

The double Darboux method appears to be a quite general and useful method
to generate new sets of quantum cosmological solutions for the
one-dimensional cases. This is so because
any potential in the Schr\"odinger equation has a classical continuum of
positive energy nonnormalizable solutions. Ryan and collaborators \cite{Ry}
were among the first to pay attention to the continuum part of the
Wheeler-DeWitt spectrum. Selecting by means of some
preliminary physical arguments that are of quantum scattering type
one of the continuum solutions \cite{Ry},
one can perform the double Darboux construction on that solution
and generate by this means families of
cosmological potentials as well as strictly isospectral
cosmological wavefunctions.
The quantum Taub model just illustrates this nice feature of
the double Darboux method.
The parameter $\lambda$ looks like a decoherence parameter
embodying a sort of cosmological dissipation (or damping) distance.
Its true physical origin, i.e., the right interpretation,
should be further investigated.

Finally, one can easily apply combinations of any pairs of
Abraham-Moses procedure, Pursey's one, and the Darboux one.
However, only the double Darboux method used here leads to reflection and
transmission amplitudes identical to those of the original potential.
 
\section*{Acknowledgment}
This work was partially supported by the Consejo Nacional de Ciencia y
Tecnolog\'{\i}a (CONACyT) (Mexico) Projects No. 4862-E9406 and 4868-E9406.

\end{document}